\documentclass{appolb}
\usepackage{graphicx}
  \usepackage{xcolor}
\usepackage{ifthen}
\usepackage{amsmath}
\usepackage{amssymb}
\usepackage{mathtools, nccmath}
\usepackage{graphicx, wrapfig}
\usepackage{slashed, bm}
\usepackage[utf8]{inputenc}
\usepackage{authblk}
\usepackage[normalem]{ulem}

\begin{document}
\title{Entanglement entropy of proton and its relation to thermodynamics entropy.
\thanks{Presented at Epiphany 2024}%
}
\author{Krzysztof Kutak
\address{Institute of Nuclear Physics, Polish Academy of Sciences, 
 ul.~Radzikowskiego 152, 31-342, Krak\'ow, Poland}
\\
}
\maketitle
\begin{abstract}
I discuss the thermodynamics-based derivation of the formula for the entanglement entropy of a system of gluons. The derivation is based on an approach where saturation and the Unruh effect were used to obtain and discuss the entropy of gluons. The formula agrees, in the high-energy limit, up to a numerical factor, with more recent results, where arguments based on the density matrix and bipartition of the proton were used to obtain the formula. I also discuss the relation of entropy as obtained in BFKL in DLL approximation and with the application of the BK equation. 
\end{abstract}
  
\section{Introduction}
In the arxiv:1103.3654v1 and arxiv: 1103.3654v2 versions of the paper \cite{Kutak:2011rb} it has been shown that if one assumes that saturation scale acts as an effective mass of a system of gluons that populate proton boosted to high rapidity and furthermore as effective temperature then one can obtain thermodynamic entropy which depends linearly on the rapidity
\begin{equation}
    S=\pi \lambda y
\label{eq:entropy1}
\end{equation}
where, $S$ is entropy of gluons  $y$ is rapidity and $\lambda$ will be introduced later.\\
The discussion leading to this formula relied on the argumentation that decelerating hadron in the color field of another hadron effectively experiences temperature in its rest frame in accord with the Unruh effect \cite{Kharzeev:2005iz}. The deceleration is of the order of the saturation scale $Q_s$ where the saturation scale signals the emergence of a dense system of gluons.
Furthermore, the motivation comes also from studies of the thermalization problem of nuclear matter where it is argued that Color Glass Condensate \cite{Gelis:2010nm} provides appropriate initial conditions for subsequent thermalization.\\
More recently however there has been substantial progress in understanding from more fundamental  principles grounded in quantum mechanics and quantum field theory the origin of entropy production in high energy collisions \cite{Peschanski:2012cw,Stoffers:2012mn,Kovner:2015hga,Kovner:2018rbf,Armesto:2019mna,Kovner:2022jqn,Berges:2017hne,Hagiwara:2017uaz,Neill:2018uqw,Ehlers:2022oke,Dumitru:2022tud,Dumitru:2023qee,Asadi:2023bat,
Gursoy:2023hge,Liu:2023eve,Liu:2022qqf,Liu:2023zno,Liu:2022ohy,Asadi:2022vbl,Dumitru:2023qee}.
We will focus on the result obtained in the  papers\cite{Kharzeev:2017qzs,Liu:2022bru,Liu:2022hto,Armesto:2019mna} where the entropy has been shown to depend linearly on rapidity. 
This behavior of entanglement entropy has been shown to be in accord with measured hadronic entropy \cite{H1:2020zpd,Hentschinski:2021aux,
Hentschinski:2022rsa,Hentschinski:2023izh,
Tu:2019ouv,Gotsman:2020bjc}.
In particular in the paper \cite{Kharzeev:2017qzs} the authors considered the Deep Inelastic Scattering process where the virtual electron probes only part of the proton's wave function and therefore introduces bi-partition of the target. This necessarily leads to the rising of entanglement of observed and unobserved degrees of freedom and therefore to entanglement entropy. 
Using the  equation that
describes the rapidity evolution of probability
for $n$ parton state $p_n(y)$  after solving and evaluating von Neuman entropy they obtain
\begin{equation}
    S(y) = \ln\left(e^{\lambda y} - 1\right) + e^{\lambda y} \ln\left(\frac{1}{1 - e^{-\lambda y}}\right)
\end{equation}
and taking the asymptotics of $y\rightarrow \infty$ they obtain the expression\footnote{The Authors of \cite{Kharzeev:2017qzs} used symbol $\Delta$ while I use $\lambda$}
\begin{equation}
    S=\lambda\,y
    \label{eq:formula2}
\end{equation}
In the 1+1 dimensional model the $\lambda$ is interpreted as the BFKL intercept and reads $\lambda=4\frac{N_c\alpha_s}{\pi}\ln 2$ while in the 3+1 dimensional case it reads $\lambda = \frac{N_c \alpha_s}{\pi} \ln(r^2 Q_s^2)$ where $r$ is the size of the dipole. 
The similar structure was also obtained within $3+1$ dimensional dipole model in the double logarithmic approximation \cite{Liu:2022hto,Liu:2022bru}.\\
The formula eq. (\ref{eq:entropy1}) is up to a constant the same as in eq. (\ref{eq:formula2}) which results from an asymptotic expansion of the complete expression (the asymptotic expansion here means that one is reaching a maximally entangled state \cite{Hentschinski:2023izh}).
This is also consistent with the thermodynamic vs. statistical-based approach where quantities tend to match after a long time passes, the role of time is played here by rapidity.
The $\lambda$ is the speed of growth of low $x$ or moderate $x$ gluons.
\section*{Entropy formula}
One can reconcile eq. (\ref{eq:entropy1}) with eq. (\ref{eq:formula2}) by rescaling $\lambda $ in the equation that connects the saturation scale with temperature, eq. (\ref{eq:entropy1}), through the introduction of a constant factor $c = \pi$, as expressed by
\begin{equation}
    T=\frac{c\,Q_s}{2\pi}
    \label{eq:unruh}
\end{equation}
While this is arbitrary one should keep in mind that the eq. (\ref{eq:unruh}) with $c=1$ is based on qualitative arguments that the deceleration is equal to the saturation scale. Because of that the formula the formula (\ref{eq:entropy1}) is approximate. 
The more fundamental derivation of (\ref{eq:entropy1}) is presented in \cite{Kharzeev:2017qzs,Liu:2022bru,Liu:2022hto}. 
Now we use the thermodynamic relation between energy and entropy:
\begin{equation}
dE=TdS
\label{ref:termo}
\end{equation}
and set $dE=dM$
gives:
\begin{equation}
dM=TdS
\label{eq:termo}
\end{equation}
Using the argument that the saturation scale acts as an effective mass of a system of gluons we have
\begin{equation}
dM=dQ_s(x)
\end{equation}
In the next step, we use eq.(\ref{eq:unruh}) which allows us to link the saturation scale to entropy:
\begin{equation}
\frac{dQ_s(x)}{Q_s(x)}=c \frac{ dS}{2\pi}
\end{equation}
which leads to:
\begin{equation}
S=\frac{\pi}{c}\ln (Q_s^2(x)/Q_0^2)
\label{eq:entropia}
\end{equation}
 and we set the lowest entropy state to zero.
Now using that saturation scale is approximately $Q_s^2=Q_0^2(x_0/x)^\lambda$ \cite{Golec-Biernat:1998zce}
and define rapidity as $ y=\ln(x_0/x)$ ($x_0$ and $Q_0$ are constants) we obtain:
\begin{equation}
S=\,\,\lambda\,y
\label{eq:disentropy}
\end{equation}
where the integration constants have been chosen to match the formulas. The basic observation that allowed the derivation of this formula within the thermodynamic approach is that the saturated system of gluons is characterized by only one scale the saturation scale $Q_s$.
This feature can be used to express the entropy formula in terms of a number of gluons in analogy to \cite{Kharzeev:2017qzs} (see also discussion along this lines in \cite{Kutak:2011rb}).
We will use the GBW gluon density that reads
\begin{equation}
    \mathcal{F}(x, k^2) =\frac{N_c S_\perp}{\alpha_s 8\pi^2}\frac{k^2}{Q_s^2} e^{-\frac{k^2}{Q_s^2}}
\end{equation}
After integrating over $k^2$ we obtain 
\begin{equation}
    xg(x)=\int^{\infty}_0 dk^2 \mathcal{F}(x,k^2) = \frac{N_c S_\perp}{\alpha_s 8\pi^2}Q_s^2
    \label{eq:gluonintegrated}
\end{equation}
Using (\ref{eq:entropia}) and (\ref{eq:gluonintegrated}) that we may write
\begin{equation}
\label{eq:entropyglue}
    S=\ln{xg(x)}+const
    \end{equation}
    where the constant can be absorbed in the $xg(x)$. 
The expression above was obtained assuming a specific form of unintegrated gluon density. However, the crucial point is that we work in saturation-dominated regions of phase space. One could use any other low $x$  dipole gluon density with saturation as they behave as ${\cal F}\sim k^2$ and integrate it up to saturation scale to arrive at a result that would differ by a constant.
Another derivation of the equation (\ref{eq:entropyglue}) was obtained in double leading logarithmic approach (DLL). The derivation allowed to account for hard scale dependence \cite{Liu:2022bru}.\\ 
Some similarities of behaviour of entropy obtained within DLL and saturation based approximations can be understood better with the help of momentum space versions of Balitsky-Fadin-Kuraev- Lipatov \cite{Balitsky:1978ic} and Balitsky-Kovchegov \cite{Balitsky:1995ub,Kovchegov:1999yj} evolution equations. 
 As it is well known the BFKL equation for unintegrated gluon density ${\cal F}(x,k^2)$ 
 \begin{equation}
{\cal F}(x,k) = {\cal F}^{(0)}(x,k)  +
 \overline{\alpha}_s \int_x^1 \frac{dz}{z} \int {dk'^2} 
\bigg[ \frac{{\cal F}(\frac{x}{z},k')}{|k^2-k'^2|}-\frac{k^2}{k'^2}\frac{{\cal F}(\frac{x}{z},k)}{|k^2-k'^2|}
+\frac{k^2}{k'^2}\frac{{\cal F}(\frac{x}{z},k)}{\sqrt{k^4+4k'^4}}
\bigg] .
    \label{eq:virtual_kms_F}
\end{equation}
is infrared sensitive because of the presence of the anticollinear pole i.e. configurations where $k^{\prime 2} \gg k^2$ and unordered emissions in the transverse momentum. 
The equation can be solved in diffusive approximation which is far from both collinear and anticollinear region but the resulting solution is not in accord with KNO scaling found in \cite{Liu:2022bru}.\\
The BK equation which accounts for recombination of gluons and therefore models saturation has this feature that the triple pomeron vertex is dominated by the anticollinear pole which as evolution progresses is subtracted from the BFKL kernel therefore overall its contribution diminishes. This can be seen from the structure of integrals in the BK equation as shown below
\cite{Kutak:2003bd,Bartels:2007dm} (see also \cite{Motyka:2023pmt}).
\begin{multline}
{\cal F}(x,k^2) = {\cal F}^{(0)}(x,k^2)  
+  \overline{\alpha}_s \int_x^1 \frac{dz}{z} \int {dk'^2} 
\bigg[ \frac{{\cal F}(\frac{x}{z},k'^2)}{|k^2-k'^2|}-\frac{k^2}{k'^2}\frac{{\cal F}(\frac{x}{z},k^2)}{|k^2-k'^2|}
+\frac{k^2}{k'^2}\frac{{\cal F}(\frac{x}{z},k^2)}{\sqrt{k^4+4k'^4}}
\bigg]
 + \\
 -\frac{2\overline{\alpha}_s^2\pi^3}{N_c^2 R^2} \int_{x}^1\frac{dz}{z}\Bigg\{
\bigg[\int_{k^2}^{\infty}\frac{dk'^2}{k'^2}{\cal F}(x/z,k'^2)\bigg]^2 
+\; {\cal F}(x/z,k^2)\int_{k^2}^{\infty}\frac{dk'^2}{k'^2}\ln\left(\frac{k'^2}{k^2}\right) {\cal F}(x/z,k'^2)
\Bigg\}\;
\label{eq:virtual_kms_F}
\end{multline}
As one can see the integral over $k'^2$ in the nonlinear part has the lower limit set by $k^2$.
Furthermore, the diffusion behavior of the linear part of the equation is tamed by the nonlinearity \cite{Golec-Biernat:2001dqn}. 
To some extent, such features can be mimicked by the double leading logarithmic approximation of the BFKL equation where the anticollinear pole is neglected.
Furthermore, this approximation gives gluon density far from the diffusive region.

\begin{equation}
{\cal F}(x,k^2) = {\cal F}^{(0)}(x,k^2)
+ \overline{\alpha}_s \int_x^1 \frac{dz}{z} \int^{k^2}_{k_{min}^2} {dk'^2} 
\frac{{\cal F}(\frac{x}{z},k'^2)}{k^2}
\; .
    \label{eq:virtual_kms_F}
\end{equation}
We expect that the mentioned above similarities of BK and BFKL in DLL approximation might be lead to similar mechanism for the generation of entropy (at least in some region) as both of the equations have limited phase space as compared to the BFKL evolution.
However, the amount of entropy will be different as the nonlinearity in the BK equation starts to play a role and to constrain phase space when $x$ is very small and $k_T$ is small while the phase space in the DLL approximated BFKL equation is constrained from the beginning. Eventually, the BK will lead to vanishing entropy while the entropy in DLL will grow.

\section*{Conclusions}
In the paper, we revisited the thermodynamics-based derivation of the entanglement entropy formula. The formula agrees in functional form with the asymptotic limit of the expression obtained by using the dipole cascade model \cite{Kharzeev:2017qzs}. By appropriately matching numerical factors, the formulas can be made to take the same form.
The findings of this paper demonstrate that in QCD, one can, in principle, calculate the same quantity using both a thermodynamic and a fine-grained quantum theory-based approach.
This stands in contrast to the current state of black hole physics, where calculating the entropy of black holes in a 3+1 D case within a realistic theory remains a significant challenge.
From this perspective, QCD may play a role in testing ideas for a better understanding of quantum gravity problems (through various mappings between QCD and gravity \cite{SabioVera:2014mkb}), as it has regimes in which it is nearly classical and by construction unitary. Questions along these lines and concrete ideas were formulated in \cite{Dvali:2021ooc}.

\section*{Acknowledgments}
\label{sec:Ack}

KK acknowledges
the European Union’s Horizon 2020 research and innovation programme under grant agreement No. 824093.

\appendix

\providecommand{\href}[2]{#2}\begingroup\raggedright\endgroup

\end{document}